\documentclass[journal=jctcce,manuscript=article]{achemso}

\usepackage[fleqn]{amsmath}
\usepackage{amssymb}
\usepackage{stmaryrd}
\usepackage{bm}
\usepackage{subfigure}
\usepackage[colorlinks]{hyperref}
\usepackage[colorinlistoftodos]{todonotes}
\usepackage{rotating}

\usepackage{verbatim}
\usepackage{color}

\usepackage[version=3]{mhchem} 

\newcommand{\deh}{$\Delta E_{\mathrm{H}_2}$}

\author{Geng Dong}
\affiliation{Department of Theoretical Chemistry, Lund University, Chemical Centre, P. O. Box 124, SE-221 00 Lund, Sweden}
\author{Ulf Ryde}
\email{Ulf.Ryde@teokem.lu.se}
\affiliation{Department of Theoretical Chemistry, Lund University, Chemical Centre, P. O. Box 124, SE-221 00 Lund, Sweden}
\author{Hans J{\o}rgen {Aa}. Jensen}
\affiliation{Department of Physics, Chemistry and Pharmacy, University of Southern Denmark, DK-5230 Odense M, Denmark}
\author{Erik D. Hedeg\aa rd}
\email{erik.hedegard@teokem.lu.se}
\affiliation{Department of Theoretical Chemistry, Lund University, Chemical Centre, P. O. Box 124, SE-221 00 Lund, Sweden}

\title[]
{Exploration of \ce{H2} binding to the [NiFe]-hydrogenase active site with multiconfigurational density functional theory }

\begin{document}
\begin{abstract}
The combination of density functional theory (DFT) with a multiconfigurational wave function is an efficient way to include dynamical correlation in calculations with multiconfiguration self-consistent field wave functions. These methods can potentially be employed to elucidate reaction mechanisms in bio-inorganic chemistry, where many other methods become either too computationally expensive or too inaccurate. In this paper, a complete active space (CAS) short-range DFT (CAS--srDFT) hybrid was employed to investigate a bio-inorganic system, namely \ce{H2} binding to the active site of [NiFe] hydrogenase. This system was  previously investigated with coupled-cluster (CC) and multiconfigurational methods in form of cumulant-approximated second-order perturbation theory, based on the density matrix renormalization group (DMRG).
 We find that it is more favorable for \ce{H2} to bind to Ni than to Fe, in agreement with previous CC and DMRG calculations. The accuracy of CAS--srDFT is comparable to both CC and DMRG, despite that much smaller active spaces were employed. This enhanced efficiency at smaller active spaces shows that  CAS--srDFT can become a useful method for bio-inorganic chemistry. 

\end{abstract}


\section{Introduction}

Quantum mechanical (QM) methods today play a prominent role in many branches of chemical science. In particular, Kohn--Sham density functional theory (DFT) has made a large impact owing to its computational efficiency and often accurate results. \cite{burke2012,harvey2006,marques2004,Casida_tddft_review_2012,Kirchner2007} However, for systems with dense frontier orbital manifolds and with degenerate or near-degenerate electronic states, DFT can be inaccurate, which is often seen for transition-metal complexes in biological systems.\cite{pierloot2011}. Thus, methods that can handle such cases are needed. The coupled cluster (CC) methods can be highly accurate, but they may also deteriorate for multiconfigurational systems and are considerably more expensive, if at all feasible. 
The alternative is to employ a multiconfigurational wave function.   
One of the most common multiconfigurational methods is the complete active space (CAS) approach, in which the orbitals are divided into active and inactive spaces. Within the active space, all configurations are included in a full configuration interaction (full-CI) calculation, thus incorporating any multiconfigurational character. Combining the CAS with a self-consistent field (SCF) procedure leads to  the complete active space self-consistent field (CASSCF) method.\cite{ruedenberg1979,roos1980b,siegbahn1981,jensen1984,olsen1983,jensen1986} On the one hand, the accuracy of CAS-based methods depends on the size of the active space, in which all important orbitals should be included. On the other hand, the computational effort also rises steeply with the size of the active space so that traditional CAS implementations are restricted to about 16--18 orbitals. This puts limitations to what type of systems that can be studied; for instance, systems with two transition metals are normally already too large. 
Methods that allow more orbitals in the active space have been introduced in recent years, for example the density matrix renormalization group (DMRG) method\cite{white1992,chan2004,marti2011,wouters2014b,legeza2003,keller2015,yanai2014,freitag2017}. 

Another serious problem is that all CAS methods, even with very large active spaces, neglect a major part of the dynamical correlation. 
To recover the missing dynamical correlation, perturbation theory is normally employed after a CASSCF or DMRG--SCF calculation,  as done in  CASPT2\cite{andersson1990,andersson1992,wouters2014b,kurashige2011,kurashige2014} or NEVPT2\cite{angeli2001,guo2016,freitag2017}. However, the perturbation correction comes with additional high computational cost.

An efficient method to recover the dynamical correlation in multiconfigurational methods is to merge DFT with a multiconfigurational wave function, thereby capitalizing on the efficient treatment of semi-local dynamical electron correlation within DFT methods. Simultaneously, such a hybrid method has the advantage that the multiconfigurational wave function can include static correlation. \cite{grimme1999,marian2008,manni2014,savin1995,fromager2007} 
In this paper, we explore the multiconfigurational short-range DFT (MC--srDFT) method. It exploits the concept of range separation of the two-electron repulsion operator to merge DFT with a multiconfigurational wave function. With a recent extension of the MC--srDFT method to a polarizable embedding framework\cite{olsen2010,hedegaard2015,hedegaard2016b}, the method can also be employed on biological systems, and  the method may be a promising approach to use for metalloenzymes.  However, 
the MC--srDFT method has mostly been benchmarked for $s$- and $p$-block atoms, diatomic molecules\cite{savin1995,savinbook,angyan2005,fromager2008,fromager2010,goll2005,fromager2013}, and organic systems\cite{hubert2016a,hubert2016b,hedegaard2013a,hedegaard2016a}. Studies of transition metals are more rare\cite{fromager2013,olsen2017}. Before addressing full enzymes, we first need to ensure that the results of MC--srDFT are in agreement with previous accurate calculations for biologically relevant cases and this is the purpose of the present paper.   

We investigate 
the binding of H$_2$ to the active site of [NiFe] hydrogenase,
for which previous studies have given ambiguous results. On the one hand, experimental
studies with CO or Xe gas-diffusion have predicted that \ce{H2} binds to  Ni.\cite{ogata2002,montet1997,volbeda2003} On the other hand, Fe is the expected binding site from the organometallic perspective\cite{bookh2ase1,bookh2ase2}.
Various DFT studies have predicted that H$_2$ binds to Ni or to Fe with the active site in the Ni(II) singlet, or even to Fe in the triplet state.\cite{siegbahn2007,bruschi2014,niu1999,pavlov1999,wu2008,jayapal2008,kaliakin2015}  We have recently investigated the H$_2$ binding site by using CCSD(T) and cumulant approximated DMRG--CASPT2 methods\cite{geng2017}, as well as DFT-based calculations with the big-QM approach\cite{hu2013}, using 819 atoms in the QM region. In this study, we compare results obtained with the MC--srPBE method with the previous CCSD(T) and DMRG--CASPT2 results, and show that MC--srDFT comes to the same conclusions. We furthermore study the method's dependence on the size of the active space and the employed basis set. 
 
\section{Computational method} \label{compmet}

\subsection{The MC--srDFT method}

The MC--srDFT method is a hybrid between wave function theory (WFT) and density functional theory (DFT). 
The method relies on range-separation of the two-electron 
repulsion operator into long-range and short-range parts\cite{savin1995,iikura2001} 
\begin{equation}\hat{g}_{\rm ee}(1,2)= \hat{g}^{\rm lr}_{\rm ee}(1,2) + \hat{g}^{\rm sr}_{\rm ee}(1,2) . 
\label{sr-DFTCoulpart}
\end{equation}
Several forms of the range-separated operators have been suggested\cite{savin1995,savinbook,erferfgaufunc}. 
We use in this work a range-separation operator based on the error function\cite{leininger1997,pollet2002,fromager2007,fromager2010} 
\begin{equation}
  \hat{g}^{\rm lr}_{\rm ee}(1,2) = \frac{\text{erf} (\mu | \textbf{r}_{1} - \textbf{r}_{2} |)}{|\textbf{r}_{1} - \textbf{r}_{2}|}; \qquad
  \hat{g}^{\rm sr}_{\rm ee}(1,2) = \frac{1- \text{erf} (\mu | \textbf{r}_{1} - \textbf{r}_{2} |)}{|\textbf{r}_{1} - \textbf{r}_{2}|}   \label{lrpart}  ,
\end{equation}
where $\mu$ is the range-separation parameter, measured in bohr$^{-1}$ in this article. This parameter is to some degree adjustable and slightly different values have been employed in the literature (we discuss this point further below). In limiting cases, a value of $\mu=\infty$ reduces MC--srDFT to multiconfigurational SCF (MCSCF), a pure wave function method, whereas $\mu=0$ reduces MC--srDFT to a pure Kohn-Sham DFT method. Both $g^{\text{lr}}_{\rm ee}(1,2)$ and $  g^{\text{sr}}_{\rm ee}(1,2)$ depend on the choice of $\mu$, but this dependence has been left out in all equations for brevity,
because $\mu$ is selected {\it a priori} and then kept fixed.
The effective electronic Hamiltonian employed in MC--srDFT is
\begin{equation}\label{srKShamil}
 \hat{H}[\rho] = \hat{h} + \hat{V}^{{\rm sr}}_{\text{Hxc}}[\rho] + \hat{g}^{\text{lr}}_{\rm ee} ,  
\end{equation}
where $\hat{h}$ contains the usual one-electron operators (kinetic energy and nuclear-electron attraction),
$\hat{g}^{\text{lr}}_{\rm ee}$ was defined in Eq.~\eqref{lrpart}, and the short-range DFT potential operator is defined through (see e.g.~ref.~\citenum{fromager2008}) 
\begin{equation}
\hat{V}^{{\rm{sr}}}_{\text{Hxc}}[\rho] =  \int \text{d}\textbf{r}\, v^{{\rm sr}}_{\rm Hxc}[\rho]\hat{\rho}(\textbf{r}) .
\end{equation}
Here $\hat{\rho}(\bm{r})$ is the density operator and $v^{{\rm sr},\mu}_{\rm Hxc}$ is the short-range adapted, $\mu$-dependent Hartree exchange--correlation potential 
\begin{equation}
 v^{{\rm sr}}_{\rm Hxc}[\rho] = \frac{\delta E^{{\rm sr}}_{\rm Hxc}}{\delta \rho(\textbf{r})} . 
\end{equation}
 It should be stressed that special exchange--correlation functionals are a prerequisite for range-separated wave function DFT hybrids (this point is explained 
thoroughly in ref.~\citenum{fromager2015}). We use 
in this work the short-range PBE-based srPBE functional by Goll {\it et al.}\cite{goll2005,goll2006}. In all cases, the applied multiconfigurational wave function ansatz was of the CASSCF type. Further,   in a few trial calculations (reported in the SI), we also employed a wave function \textit{ansatz} based on M{\o}ller--Plesset second order perturbation theory (MP2). 
Since the applied multiconfigurational wave function ansatz was of the CASSCF type, we will henceforth refer to MC--srDFT with respect to the choices of multiconfigurational wave function and functional, i.e., CAS--srPBE for the method employed in this paper, and MC--srDFT for the general method.

For the range-separation parameter, most studies on range-separated DFT hybrids\cite{vydrov2006,baer2010} employ values between 0.33--0.5 bohr$^{-1}$. For MC--srDFT, a value of $\mu=0.4$ bohr$^{-1}$ has been suggested based on natural occupation numbers and differences between HF--srDFT and CAS--srDFT ground-state energies of small organic systems.\cite{fromager2007}. Benchmark studies on excitation energies \cite{hubert2016a,hubert2016b,hedegaard2016a} for organic systems have confirmed that this value provides accurate results.   
Using both MP2--srPBE and CAS--srPBE models, we have tested range of $\mu$\ values (see the supporting information, Table S1). These results show that  $\mu$ values between 0.5 and 0.3 gives relative energies of \textbf{H$_2$-Fe} and \textbf{H$_2$-Ni} close to the energies obtained with DMRG and CCSD(T). Since $\mu=0.4$ is both accurate and consistent with previous suggestions, we here employ  $\mu=0.4$ bohr$^{-1}$.  
 
All calculations were carried out with a development version of the DALTON program.\cite{DaltonDevel,WIREdalton} Further details about the MC--srDFT method, as well as the implementation, can be found elsewhere\cite{pedersenphd2004}.

\subsection{Model systems and basis sets}

As the name indicates, the active site of [NiFe] hydrogenase consists of a Ni ion and a Fe ion. The former is coordinated to four Cys residues, two of which are also bridging to the Fe ion. The latter also coordinates one CO and two \ce{CN-} ligands.  In this paper, we compare the stability of two binding modes of \ce{H2} to this site, viz. binding side-on to Ni or to Fe. The two binding modes will be called \textbf{H$_2$--Ni} and \textbf{H$_2$--Fe}, and they are shown in Figure 1 (note that \ce{H2} actually bridges the two metal ions in the \textbf{H$_2$--Fe} binding mode). In analogy with our previous study\cite{geng2017}, we used for each state three models of increasing size, also shown in Figure 1. In the smallest model 1, the four cysteine ligands were modeled by \ce{HS-} groups, whereas in the other two models they were modeled by \ce{CH3S-}. In the largest model 3, two second-sphere residues were included, Glu34 and His88 (residue numbering according to the crystal structure with PDB entry 1H2R\cite{higuchi1999}), modeled by acetic acid and imidazole, respectively. 
The structures were taken from our previous study\cite{geng2017} and were optimized with the combined quantum mechanics and molecular mechanics (QM/MM) approach at the TPSS/def2-SV(P) level of theory \cite{ryde2001,ryde1996,tao2003,schafer1994}  in the singlet state.
Thus, both the Ni and Fe ions are in the low-spin +II oxidation state, corresponding to the spectroscopic Ni-SI$_a$ state of [NiFe] hydrogenase\cite{lubitz2014}. 
  
\graphicspath{ {/Users/geng/Desktop/srdft/FIGURES/} }
\begin{figure}
\centering
\includegraphics[width=0.95\textwidth]{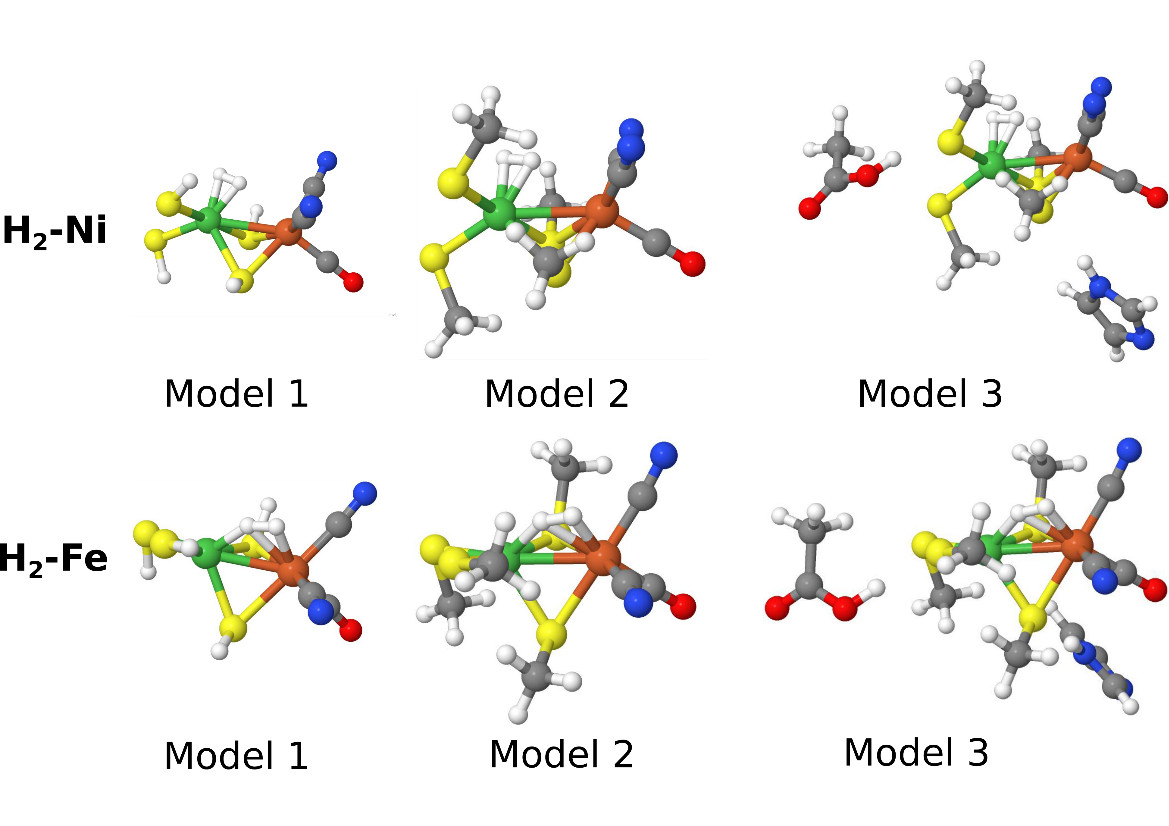}
\caption{The models we used in this work. \label{systems} }
\end{figure}

The calculations presented here were carried out with three basis sets of increasing size, denoted B1--B3. For the smallest one (named B1), the cc-pVTZ \cite{bs890115d,bs920501kdh} basis set was employed for the Ni and Fe ions, and the cc-pVDZ \cite{bs890115d} basis set was used for the other atoms. 
The effect of increasing the basis set was investigated by using the cc-pVTZ basis set on all atoms except H, for which the cc-pVDZ basis set was used. This basis set resembles the ANO-type basis set employed in Ref.~\citenum{geng2017} and is denoted B2. In addition, we have added a calculation  with a basis set similar to B2, but in which the \ce{H2} molecule bound to Ni or Fe is also described with the cc-pVTZ basis set. Thus, the important \ce{H2}-molecule is (in contrast to ref.~\citenum{geng2017}) also described with a triple-zeta basis set. We denoted this last basis set B3. The MC--srDFT method has during this study undergone development to become more efficient and this effort is ongoing. Still, a large number of inactive electrons does pose a challenge for the current implementation of MC--srDFT. Therefore the basis sets B2 and B3 were  used only for model 1.  It should be emphasized that this is not a challenge of the method itself and standard techniques (e.g. Cholesky decomposition) can straightforwardly be applied to MC--srDFT. Relativistic effects were considered by using a standard second-order Douglas--Kroll--Hess (DKH2) Hamiltonian.  \cite{douglas1974,hess1985,hess1986}

\subsection{Selection of active spaces}

The selection of active space is of highest importance for an MCSCF calculation and 
different strategies have been proposed for this selection. One strategy has been to rely on identifying orbitals from the chemical context.\cite{veryazov2011}  For transition metal complexes, this has typically led to the suggestion that all $3d$ orbitals and a few ligand orbitals should be included, and preferably also an additional (double-shell) of $3d'$ orbitals. A different strategy relies on selecting orbitals 
 based on natural occupation numbers from methods where the predicted occupation numbers are qualitatively correct. This could be either MP2\cite{jensen1988} or a computationally cheap CI method.  Typically, one would select orbitals with occupation numbers significantly different from 2 or 0. Rules for selection of active spaces are not as well established for short-range DFT methods. Occupation numbers based on MP2--srDFT have previously been discussed\cite{hedegaard2013b} and it was noted these natural occupation numbers are much closer to 2.0 and 0.0 than their MP2 counterparts. This is expected because the short-range density functional effectively includes dynamical Coulomb-hole correlation. Hence, orbitals with occupation numbers below 1.98 in MP2- or MC-srDFT can be expected to show strong correlation and orbitals with occupation numbers of around 1.98 or 0.02 should preferably be included in the active spaces.  Importantly, since we here investigate the relative energy of two species, the chosen active spaces of the two species must be comparable.  

The MP2--srPBE occupation numbers for the two complexes (model 1) are compiled in Tables S2 and S3, and our initial selection of orbitals for the CAS--srDFT calculations was based on these. Tables S2 and S3 also contain occupation numbers for a number of different $\mu$\ values, but we focus here on $\mu= 0.4$ bohr$^{-1}$. 
Occupation numbers with a similar magnitude should preferably be included as a group and we have initially selected a CAS(10,10) space, 
for which there is a clear change in occupation numbers between the selected 10 orbitals and the orbitals not included (for both \textbf{H$_2$--Ni} and  \textbf{H$_2$--Fe}).
A larger active space is more challenging to define: For  \textbf{H$_2$--Ni}, selecting CAS(12,12) or CAS(14,14) will mean including and excluding orbitals with rather similar MP2--srPBE occupation numbers.
The CAS(16,16) choice seems better, but this is rather large. On the other hand, for  \textbf{H$_2$--Fe}, CAS(10,10) or CAS(16,15) seems appropriate based on the MP2--srDFT occupation numbers.  

Considering that the  MP2--srPBE occupation numbers might not reflect  the "true" occupation number (i.e. occupation numbers obtained with a full-CI-srPBE approach), we initially investigated CAS(10,10), CAS(14,14) and CAS(16,16) for both species. The corresponding CAS--srPBE occupation numbers are also shown in Tables S2 and S3. For CAS(16,16), we start to include orbitals with either very high or very low occupation numbers (above 1.99 or below 0.01), which affects the convergence. The CAS(16,16)--srPBE calculation for \textbf{H$_2$--Fe} also shows that the occupation numbers for the last two orbitals in what would correspond to a CAS(12,12) become even closer than for the MP2--srPBE calculation. Hence, CAS(12,12) will become unstable and prone to get stuck in local minima, which was confirmed by a trial calculation with this active space. The orbitals causing these difficulties are involved in the \ce{Fe-CN} and \ce{Fe-CO} bonds, and care must be taken to include these orbitals uniformly in the two states. This is done in  CAS(14,14), which is the largest active space that can be considered balanced (and it is also feasible for the larger models 2 and 3).

\begin{figure}
\centering
\includegraphics[width=1.1\textwidth]{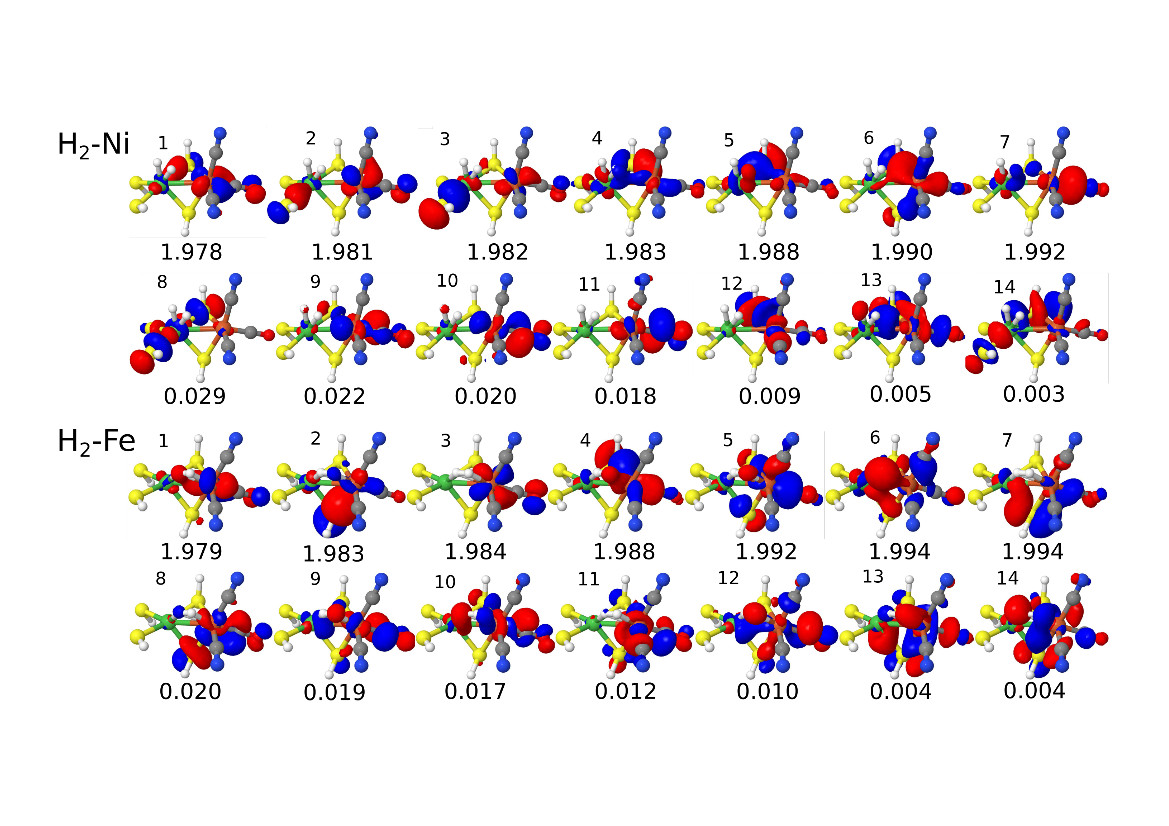}
\caption{ Active natural orbitals and their occupation numbers from the CAS(14,14)--srPBE calculation on model 1. \label{cas-14-14-orb}}
\end{figure}

  Visual inspection of the CAS(14,14) orbitals in Figure \ref{cas-14-14-orb} shows that this active space includes the Ni $3d$-orbitals, the \ce{H2} and metal--ligand (CO $\pi$-type) orbitals, although the orbitals are more delocalized than the pure DMRG-SCF (or CASSCF) orbitals in refs.~\citenum{delcey2014,geng2017}. Further reduction of the active space to CAS(10,10), leads to exclusion of orbitals that are partly on hydrogen and the Ni ion, and we therefore prefer to include these two orbitals (i.e.~orbitals 4, 5, 13 and 14 are included for \textbf{H$_2$--Ni} in Figure \ref{cas-14-14-orb}, compared to Figure S2). Furthermore, the occupation number of the Ni orbital in  \textbf{H$_2$--Ni} (orbital 4 in Figure \ref{cas-14-14-orb}) is around 1.98 in the CAS(14,14) calculations and thus rather close to two of the other orbitals in the active space. This indicates that this orbital should be included. 
  
  Expanding the calculations to CAS(16,16), introduces orbitals that are mainly on bridging sulfur atoms and can be considered less important.  For instance,  for \textbf{H$_2$-Ni}, the additional orbitals compared to CAS(14,14) are orbitals 8 and 14 in Figure S5. Although we here focus on  CAS(14,14), it should be noted that the effect on the calculated (relative) energies is in fact small (2 kJ/mol and below), as will be discussed in next section. For models 2 and 3, we also focus on CAS(14,14), but we have employed both CAS(10,10) and CAS(14,14) active spaces to probe the effect of the active spaces for these larger models as well. he corresponding active space orbitals are shown in the SI. 
  Finally, we note that we also attempted to select orbitals based on calculations with larger $\mu$\ values, but this procedure was less satisfactory (shortly discussed in the SI).

\graphicspath{ {/Users/geng/Desktop/srdft/FIGURES/14in14/} }



\section{Results and Discussion}
In this study, we have compared the results of CAS--srPBE with previously published CCSD(T) and cumulant approximated DMRG--CASPT2 calculations for the two binding modes of \ce{H2} to the active site of [NiFe] hydrogenase.\cite{geng2017} We discuss first the smallest model (model 1) and then the two larger models (models 2 and 3) in separate sections.

\subsection{Calculations with model 1}

The energy difference between the \textbf{H$_2$--Ni} and \textbf{H$_2$--Fe} states (\deh) calculated with the CAS--srPBE method is compared to previous CCSD(T) and DMRG--CASPT2 results in Table 1. We report CAS--srPBE results with the CAS(10,10), CAS(14,14) and CAS(16,16) active spaces.
In all cases, CAS--srPBE predicts that the \textbf{H$_2$--Ni} state is most stable, in agreement with the CCSD(T) and DMRG--CASPT2 results. The effect of expanding the active space from CAS(10,10)--srPBE to CAS(14,14)--srPBE is only 2 kJ/mol (see the Methods section for a description of the orbitals within the two active spaces). The CAS(10,10)--srPBE predicts that \textbf{H$_2$--Ni} is 15 kJ/mol more stable, whereas the difference with CAS(14,14)--srPBE is 17 kJ/mol. For the largest active space, CAS(16,16), the obtained energy difference changes by only 0.2 kJ/mol. Hence, there is little effect on the relative energies when expanding the active space and the CAS(14,14) active space seems to be sufficiently large for the systems studied here.  

\begin{table}[]
\centering
\caption{The energy difference ($\Delta E_{\mathrm{H}_2} = E$(\textbf{H$_2$--Fe})$ - E$(\textbf{H$_2$--Ni}) in kJ/mol) between the \textbf{H$_2$--Ni} and \textbf{H$_2$--Fe} states (model 1) calculated with the CAS--srPBE method, and compared to previous CCSD(T), DMRG--CASPT2, and DFT calculations.  \label{model1}}
\begin{tabular}{lll}
\hline
Method                                & Basis      &  $\Delta E_{H2}$ \\ 
 \hline
CAS(10,10)--srPBE                     & B1         &  15.1  \\
CAS(14,14)--srPBE                     & B1         &  17.0  \\
CAS(16,16)--srPBE                     & B1         &  16.8  \\
CAS(14,14)--srPBE                     & B2         &  15.2  \\
CAS(14,14)--srPBE                     & B3         &  13.9  \\
DMRG(22,22)--CASPT2\cite{geng2017}     & ANO-RCC    &  17.7 \\
DMRG(22,22)--CASPT2$^a$\cite{geng2017}     & ANO-RCC    &  11.9  \\
CCSD(T)\cite{geng2017}                & ANO-RCC    &  18.1  \\
TPSS\cite{geng2017}                   & def2-QZVPD&  25.6  \\  \hline \hline
\multicolumn{2}{l}{$^a$ With 3s,3p correlation obtained from CCSD(T).} 
\end{tabular}
\end{table}

The CAS(14,14)--srPBE results with both the B1 and B2 basis sets (17 and 15 kJ/mol) are in good agreement with the result obtained with DMRG(22,22)--CASPT2 (18 kJ/mol), although those calculations employed a significantly larger active space. Hence, the treatment of semi-local, dynamical correlation by the srDFT part allows for the use of significantly smaller active spaces compared to traditional MR methods.
 It should be noted that the CAS--srPBE calculations also show a rather modest basis set dependence. The basis sets increase from B1 to B2 only lowers the obtained energy-difference by 2 kJ/mol. Further increasing it to B3 lowers the energy-difference by another 1 kJ/mol, yielding a final result of 14 kJ/mol. 
 
 At this point we emphasize that recent studies have noted that multireference perturbation theory to second order does not always recover the 3s,3p correlation well.\cite{pierloot2017} Table \ref{model1} also reports a DMRG(22,22)--CASPT2 result obtained without 3s,3p correlation, but including an estimate of this semi-core correlation from CCSD(T). The resulting energy difference was then 12 kJ/mol.\cite{geng2017} Thus, our best CAS(14,14)-srPBE result (14 kJ/mol) is within 2 kJ/mol of this corrected DMRG--CASPT2 value, and within 4 kJ/mol of the CCSD(T) result.  From the above discussion, we can thus conclude that both CAS(14,14)--srPBE and DMRG(22,22)--CASPT2 reproduce the CCSD(T) data well.


\subsection{Calculations with models 2 and 3}

Next, we carried out CAS--srPBE calculations also for the two larger models 2 and 3 in Figure \ref{systems}. 
The results are shown in Table \ref{res_model_2_and_3} and are compared to the corresponding results obtained with DMRG--CASPT2. It can be seen that the two approaches give similar trends: The energy differences increase in model 2 (24--39 kJ/mol), whereas inclusion of models of two nearby amino-acids counteracts this increase, so that in model 3, the energy difference  decreases again to 8--19 kJ/mol. 
\begin{table}[]
\centering
\caption{The \deh\ energy difference (in kJ/mol) between the \textbf{H$_2$--Ni} and \textbf{H$_2$--Fe} states calculated with the CAS--srPBE method, compared to previous DMRG--CASPT2 and DFT calculations for models 2 and 3. }
\label{res_model_2_and_3}
\begin{tabular}{llrr}
\hline
Method                                 & Basis      & \multicolumn{2}{c}{\deh} \\ 
                                       &            & Model 2 & Model 3 \\ \hline
CAS(10,10)--srPBE                      & B1         & 24.3    & 7.9     \\ 
CAS(14,14)--srPBE                      & B1         & 27.7    & 10.5    \\
DMRG(22,22)--CASPT2\cite{geng2017}     & ANO-RCC    & 37.7    & 15.2    \\
DMRG(22,22)--CASPT2$^a$\cite{geng2017} & ANO-RCC    & 33.0    & 11.1    \\
DMRG(22,22)--CASPT2$^b$\cite{geng2017} & ANO-RCC    & 39.2    & 17.3    \\
TPSS\cite{geng2017}                    & def2-QZVPD & 34.0    & 18.5    \\ \hline \hline   \end{tabular}
\\
$^a$ With 3s,3p correlation obtained from CCSD(T). \\
$^b$ Extrapolated with the energy difference between CCSD(T) and DMRG(22,22)--CASPT2 for model 1, the latter with 3s,3p correlation obtained from CCSD(T).
\end{table}
For all three models, \textbf{H$_2$--Ni} is thus consistently predicted to be the most stable state and the CAS(14,14)--srPBE results are quite close to that of DMRG(22,22)--CASPT2. From Table \ref{res_model_2_and_3}, it can be seen that the differences to CAS(14,14)--srPBE are 5 and 10 kJ/mol for model 2, depending on whether the DMRG--CASPT2 included 3s,3p correlation from CCSD(T) or not. The corresponding differences for model 3 are even smaller, 1 and 5 kJ/mol. 

In Ref.~\citenum{geng2017}, it was noted that CCSD(T) was beyond the computational resources for models 2 and 3, but an estimate of a CCSD(T) result could be obtained by correcting the DMRG--CASPT2 results for models 2 and 3 with the energy-difference  between  CCSD(T) and DMRG--CASPT2 from model 1. With this correction, the results were 39 and 17 kJ/mol for models 2 and 3, respectively. Compared to these values, CAS(14,14)--srPBE underestimates the energy difference by 11 and 6 kJ/mol for models 2 and 3. Judging from the results with the smallest model, the difference is expected to decrease by 2 kJ/mol with the larger B2 basis set.  These differences are certainly acceptable and below other error sources. For instance, the protein was found to affect the energy difference by more than 25 kJ/mol in the favor of the \textbf{H$_2$--Ni} state (estimated by DFT and a 819-atom QM model).\cite{geng2017}      

Although the performance of CAS--srPBE is encouraging for applications in metalloenzymes, further improvements are possible: For instance, the accuracy of the srDFT functional can be improved. This could be achieved by either including exact (short-range) DFT exchange or by including kinetic energy dependence in the same way as in \textit{meta}-GGA functionals.\cite{goll2009}. 

\section{Conclusions}

In this paper, CAS--srPBE calculations were performed on three models of \ce{H2} bound to [NiFe]-hydrogenase. Our results indicate that \ce{H2} binding to Ni is more stable than binding to Fe, which is consistent with previous calculations with the CCSD(T) and DMRG--CASPT2 methods.\cite{geng2017} Our CAS--srPBE calculations with reduced active spaces (CAS(10,10), CAS(14,14) and CAS(16,16)) gave results close to the CAS(22,22) active space used in the previous  DMRG--CASPT2 calculations. 

For all employed model systems, the effect of extending the active space from CAS(10,10) to CAS(14,14) was found to be small (around 2 kJ/mol). For model 1, we further employed CAS(16,16), which only gave rise to a change of 0.2 kJ/mol. This is a good indication that the calculations are converged with respect to choice of active space. The effect of increasing the basis set were also quite modest:  For model 1, an increase in the basis set from cc-pVDZ to cc-pVTZ for all C, N, O and S atoms changed the CAS(14,14)--srPBE energy difference by only 3 kJ/mol. Thus, both the effect of increasing the active space and the effect of the basis set is much lower that other sources of error. For instance, a change of 7 kJ/mol was obtained by employing the 3s,3p correlation obtained from either CCSD(T) or DMRG--CASPT2 and the effect of the surrounding protein was 25 kJ/mol.  

For the larger models 2 and 3, the CAS(10,10)--srPBE and CAS(14,14)--srPBE results agree with the CCSD(T) extrapolated DMRG--CASPT2 results to within 4--9 kJ/mol. This is similar to the difference between the best DMRG(22,22)--CASPT2 and CCSD(T) results for model 1, 6 kJ/mol.
Hence, our results support MC--srDFT as a new valuable tool for bio-inorganic chemistry, with an accuracy similar to that of DMRG--CASPT2 but at a much lower computational cost (in a fully optimized implementation). The lower computational cost is achieved by means of the much smaller active spaces needed and by the replacement the perturbation correction of CASPT2 with DFT integration.

Finally, it would also be interesting to address the triplet spin-states of the two \textbf{H$_2$--Ni} and \textbf{H$_2$--Fe} intermediates. This will require extension of our current MC--srDFT implementation to functionals that depend on spin-densities, and this  development is currently ongoing. 



\acknowledgement
This investigation has been supported by grants from the Swedish research council (project 2014-5540), The Carlsberg Foundation (project CF15-0208), The European Commission, The China Scholarship Council, and COST Action CM1305 (ECOSTBio).
The computations were performed on computer resources provided by the Swedish National Infrastructure for Computing (SNIC) at Lunarc at Lund University and HPC2N at Ume\aa\  University,
and on computer resources on Abacus 2.0 provided by the Danish e-infrastructure Cooperation.


\bibliography{bib_file}

\end{document}


\beginsupplement

\maketitle

\section{Studies with different $\mu$\ values}

A study of the \deh\ energy difference between the two investigated intermediates for different $\mu$\ values was conducted, first based on MP2--srDFT calculations on model 1.
The resulting energy differences are given in Table~\ref{Mp2-srDFT}.
One notes that \deh\ decreases steadily with increasing $\mu$.
Recalling that increasing $\mu$ means moving correlation from the short-range DFT part to the long-range WFT part, a possible reason for these results could be that MP2 is inadequate to describe the energy difference in wave function theory.
We therefore decided to calculate \deh\ also with CAS(14,14)-srPBE for the most relevant $\mu$ values, and these results are also reported in Table~\ref{Mp2-srDFT}.
These values clearly support our contention that MP2 is inadequate, while CAS(14,14) is sufficiently flexible and accurate to describe the transfer of correlation effects from the DFT part to the WFT part of the MC-srDFT model when $\mu$ changes from 0.4 to 0.7.
As a side remark, the variation in the CAS(14,14) natural orbital occupation numbers in Tables \ref{MP2-occ-Ni-H2} and \ref{MP2-occ-Fe-H2} as a function of the $\mu$ value shows explicitly that the long-range correlation effects increase when $\mu$ is increased.

From these results, $\mu=$0.4--0.7 all give reasonable \deh\ energy differences compared to DMRG(22,22)-CASPT2 and CCSD(T) values (cf.~Table 1 in the paper).
Previous studies (see paper for references) on smaller systems have found that 0.4 is optimal, our results here provide no reason for advocating another value here,
and we therefore follow our earlier recommendation and use a value of $\mu=0.4$ in this study.    
\begin{table}[ht]
\centering
\caption{The \deh\ energy difference (kJ/mol) for model 1 from MP2--srPBE and CAS(14,14)-srPBE calculations with B1 and various $\mu$ values.\label{Mp2-srDFT}} 
\label{MP2-energy}
\begin{tabular}{lrrrrrrrr}
\hline
$\mu$            & 1.0 & 0.9 & 0.8 & 0.7 & 0.6 & 0.5 & 0.4 & 0.3 \\ \hline
MP2-srPBE        & 0.7 & 1.7 & 2.7 & 3.8  & 5.4  & 8.1 & 12.7 & 18.6 \\ 
CAS(14,14)-srPBE &     &     &     & 14.9 & 14.2 & 14.1 & 17.0 &    \\
\hline
\end{tabular}
\end{table}

We additionally investigated how the selection of the active space vary with different $\mu$\ values. In  Tables \ref{MP2-occ-Ni-H2} and \ref{MP2-occ-Fe-H2}, we have compiled occupation numbers of the relevant orbitals for several different $\mu$-values.  
For $\mu=1.0$, the best CAS space would be either CAS(8,8) or CAS(14,14). 
The orbitals from a CAS(14,14)--srPBE calculations are shown in Figure 5. It can be seen that metal orbitals are missing in active space and the orbitals are somewhat different from the orbitals obtained with $\mu=0.4$. This shows that care must be exercised in selecting orbitals based on different $\mu$-values.   
 
\begin{table}[ht]
\small
\centering
\caption{MP2--srPBE and CAS--srPBE natural orbital occupation numbers for \textbf{H$_2$--Ni} \label{MP2-occ-Ni-H2}}
\begin{tabular}{lllllllllll}
\\ \hline
$\mu$               &                      & \multicolumn{9}{l|}{occupation numbers}   \\ \hline
1.0 & MP2 & 1.964 & 1.964 & 1.962 & 1.961 & 1.960 & 1.959 & 1.958 & 1.957 & 1.949 \\ 
    &     & 0.046 & 0.043 & 0.042 & 0.041 & 0.033 & 0.033 & 0.030 & 0.030 & 0.026 \\ 
\hline
0.9 & MP2 & 1.966 & 1.966 & 1.966 & 1.964 & 1.964 & 1.962 & 1.961 & 1.959 & 1.950 \\ 
    &     & 0.048 & 0.040 & 0.039 & 0.037 & 0.031 & 0.031 & 0.029 & 0.027 & 0.025 \\ 
\hline                    
0.8 & MP2 & 1.971 & 1.969 & 1.969 & 1.969 & 1.966 & 1.966 & 1.965 & 1.961 & 1.952 \\ 
    &     & 0.048 & 0.037 & 0.036 & 0.033 & 0.030 & 0.029 & 0.025 & 0.023 & 0.023 \\ 
\hline                    
0.7 & MP2 & 1.974 & 1.974 & 1.972 & 1.972 & 1.970 & 1.969 & 1.968 & 1.965 & 1.957 \\ 
    &     & 0.046 & 0.033 & 0.032 & 0.029 & 0.028 & 0.027 & 0.021 & 0.020 & 0.020 \\  
    & CAS(14,14) & 2     & 2     & 1.987 & 1.978 & 1.977 & 1.966 & 1.965 & 1.964 & 1.945 \\                             
    &     & 0.055 & 0.040 & 0.036 & 0.035 & 0.022 & 0.016 & 0.013 & 0     & 0     \\  \hline                    
0.6 & MP2 & 1.979 & 1.978 & 1.977 & 1.976 & 1.974 & 1.973 & 1.971 & 1.970 & 1.964 \\ 
    &     & 0.042 & 0.028 & 0.028 & 0.026 & 0.024 & 0.023 & 0.017 & 0.017 & 0.016 \\ 
    & CAS(14,14) & 2     & 2     & 1.990 & 1.982 & 1.981 & 1.971 & 1.970 & 1.969 & 1.954 \\                              
    &     & 0.046 & 0.035 & 0.032 & 0.031 & 0.019 & 0.012 & 0.010 & 0     & 0     \\  \hline                    
0.5 & MP2 & 1.984 & 1.982 & 1.981 & 1.980 & 1.978 & 1.977 & 1.976 & 1.975 & 1.973 \\ 
    &     & 0.034 & 0.024 & 0.023 & 0.022 & 0.020 & 0.017 & 0.013 & 0.013 & 0.012 \\ 
    & CAS(14,14) & 2     & 2     & 1.992 & 1.986 & 1.985 & 1.977 & 1.975 & 1.974 & 1.966 \\                              
    &     & 0.034 & 0.028 & 0.026 & 0.026 & 0.015 & 0.008 & 0.008 & 0     & 0     \\  \hline                    
0.4 & MP2 & 1.989 & 1.987 & 1.987 & 1.986 & 1.984 & 1.983 & 1.983 & 1.982 & 1.980 \\  
    &     & 0.025 & 0.019 & 0.017 & 0.016 & 0.014 & 0.012 & 0.010 & 0.009 & 0.008 \\  
    & CAS(14,14) & 2 & 2  & 1.992 & 1.990 & 1.988 & 1.983 & 1.982 & 1.981 & 1.978 \\                           
    &     & 0.029 & 0.022 & 0.019 & 0.018 & 0.009 & 0.005 & 0.003 & 0     & 0     \\                    
    & CAS(16,16) & 2 & 1.993 & 1.991 & 1.988 & 1.986 & 1.983 & 1.982 & 1.981 & 1.977 \\                              
    &     & 0.029 & 0.024 & 0.021 & 0.019 & 0.011 & 0.006 & 0.004 & 0.003 & 0     \\  \hline
0.3 & MP2 & 1.994 & 1.993 & 1.992 & 1.992 & 1.991 & 1.990 & 1.990 & 1.989 & 1.987   \\ 
    &     & 0.016 & 0.013 & 0.010 & 0.009 & 0.008 & 0.006 & 0.006 & 0.005 & 0.005 \\   \hline  
\end{tabular}
\end{table}

\begin{table}[ht]
\small
\centering
\caption{MP2--srPBE and CAS--srPBE natural occupation numbers for  \textbf{H$_2$--Fe} \label{MP2-occ-Fe-H2}}
\begin{tabular}{lllllllllll}
\\ \hline
$\mu$               &                      & \multicolumn{9}{l|}{occupation numbers}   \\ \hline
1.0 & MP2 & 1.965 & 1.964 & 1.962 & 1.961 & 1.960 & 1.959 & 1.958 & 1.956 & 1.954 \\ 
    &     & 0.043 & 0.042 & 0.041 & 0.038 & 0.033 & 0.030 & 0.030 & 0.029 & 0.027 \\ 
\hline      
0.9 & MP2 & 1.967 & 1.967 & 1.966 & 1.964 & 1.963 & 1.963 & 1.962 & 1.958 & 1.955 \\ 
    &     & 0.041 & 0.039 & 0.038 & 0.037 & 0.031 & 0.029 & 0.028 & 0.026 & 0.026 \\
\hline  
0.8 & MP2 & 1.971 & 1.970 & 1.970 & 1.969 & 1.967 & 1.965 & 1.965 & 1.960 & 1.957 \\ 
    &     & 0.041 & 0.036 & 0.035 & 0.033 & 0.029 & 0.026 & 0.025 & 0.024 & 0.023 \\ 
\hline 
0.7 & MP2 & 1.975 & 1.974 & 1.974 & 1.973 & 1.971 & 1.969 & 1.968 & 1.964 & 1.960 \\ 
    &     & 0.041 & 0.032 & 0.031 & 0.029 & 0.027 & 0.024 & 0.021 & 0.020 & 0.019 \\
    & CAS(14,14) & 2     & 2     & 1.988 & 1.986 & 1.976 & 1.974 & 1.971 & 1.969 & 1.950 \\                               
    &     & 0.053 & 0.030 & 0.029 & 0.027 & 0.024 & 0.014 & 0.009 & 0     & 0     \\  \hline
0.6 & MP2 & 1.979 & 1.979 & 1.978 & 1.976 & 1.975 & 1.972 & 1.972 & 1.968 & 1.966 \\ 
    &     & 0.038 & 0.028 & 0.026 & 0.025 & 0.024 & 0.021 & 0.017 & 0.017 & 0.016 \\ 
    & CAS(14,14) & 2     & 2     & 1.990 & 1.988 & 1.980 & 1.978 & 1.975 & 1.972 & 1.957 \\                              
    &     & 0.046 & 0.027 & 0.025 & 0.023 & 0.020 & 0.012 & 0.007 & 0     & 0     \\  \hline
0.5 & MP2 & 1.984 & 1.984 & 1.982 & 1.981 & 1.979 & 1.978 & 1.977 & 1.975 & 1.973 \\ 
    &     & 0.033 & 0.023 & 0.021 & 0.021 & 0.019 & 0.017 & 0.013 & 0.013 & 0.012 \\ 
    & CAS(14,14) & 2     & 2     & 1.992 & 1.991 & 1.984 & 1.982 & 1.979 & 1.977 & 1.967 \\                              
    &     & 0.037 & 0.023 & 0.021 & 0.018 & 0.016 & 0.008 & 0.005 & 0     & 0     \\  \hline
0.4 & MP2 & 1.989 & 1.988 & 1.987 & 1.986 & 1.985 & 1.984 & 1.982 & 1.982 & 1.981 \\ 
    &     & 0.025 & 0.018 & 0.016 & 0.014 & 0.014 & 0.012 & 0.010 & 0.009 & 0.008 \\ 
    & CAS(14,14) & 2 & 2  & 1.994 & 1.994 & 1.993 & 1.988 & 1.984 & 1.983 & 1.979 \\                              
    &     & 0.020 & 0.019 & 0.017 & 0.012 & 0.010 & 0.004 & 0.004 & 0     & 0     \\
    & CAS(16,16) & 2 & 1.994 & 1.993 & 1.992 & 1.988 & 1.985 & 1.984 & 1.983 & 1.976 \\                            
    &     & 0.028 & 0.019 & 0.019 & 0.013 & 0.012 & 0.007 & 0.005 & 0.004 & 0       \\  \hline 
0.3 & MP2 & 1.994 & 1.993 & 1.993 & 1.992 & 1.992 & 1.991 & 1.989 & 1.989 & 1.989  \\ 
    &     & 0.016 & 0.011 & 0.010 & 0.008 & 0.007 & 0.006 & 0.006 & 0.005 & 0.005 \\   \hline    
\end{tabular}
\end{table}

\begin{figure}
\centering
\includegraphics[width=1.0\textwidth]{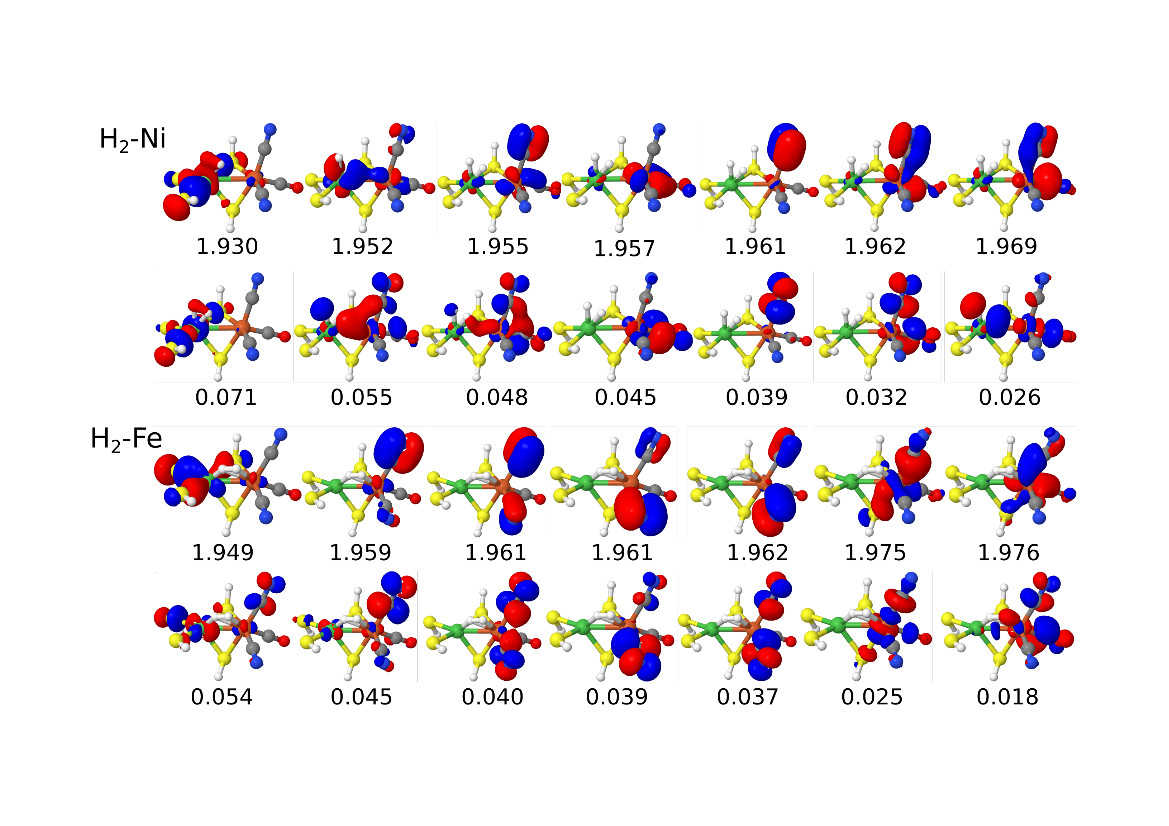}
\caption{ Active natural orbitals and their occupation numbers from a CAS(14,14)--srDFT calculation (started from the MP2 wavefunction) of model 1 with $\mu=1$}
\end{figure} 

\begin{figure}
\centering
\includegraphics[width=1.0\textwidth]{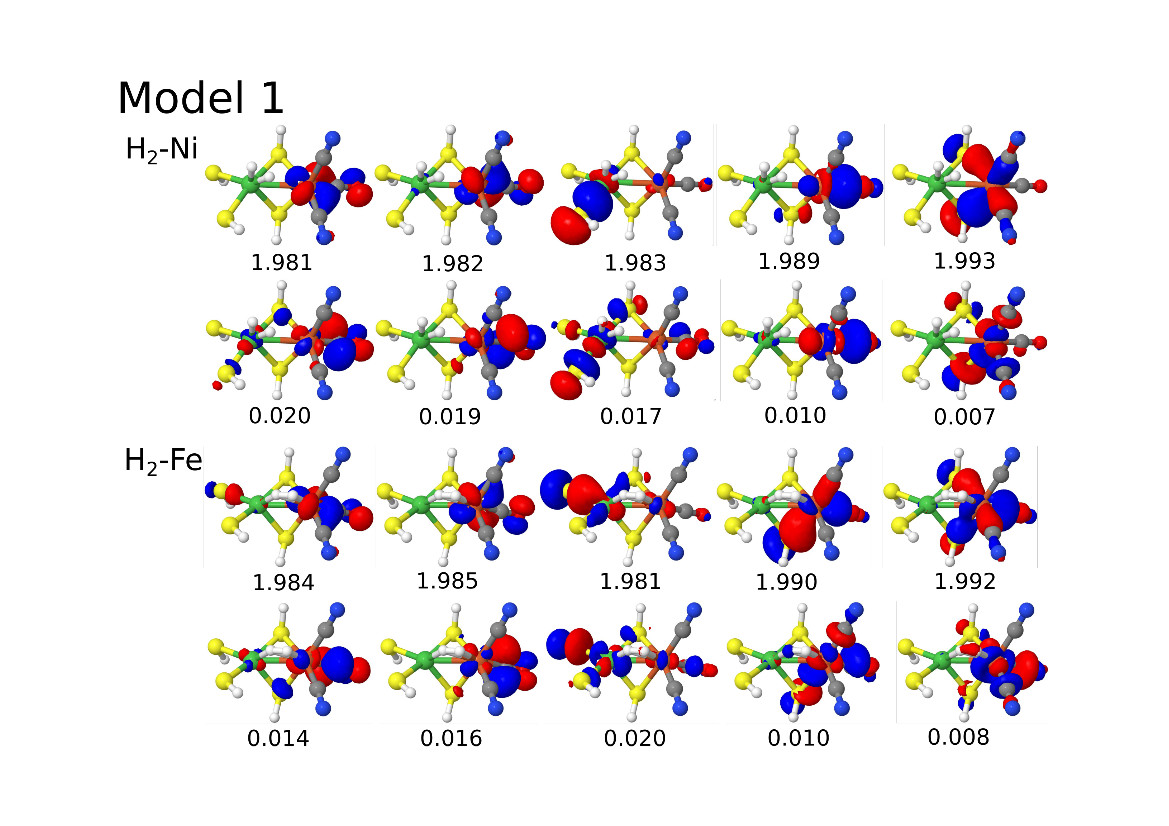}
\caption{ Active natural orbitals and their occupation numbers from a CAS(10,10)--srDFT calculation (started from the MP2 wavefunction) with $\mu=0.4$ for model 1\label{cas-10-10-orb1}} 
\end{figure} 

\begin{figure}
\centering
\includegraphics[width=1.0\textwidth]{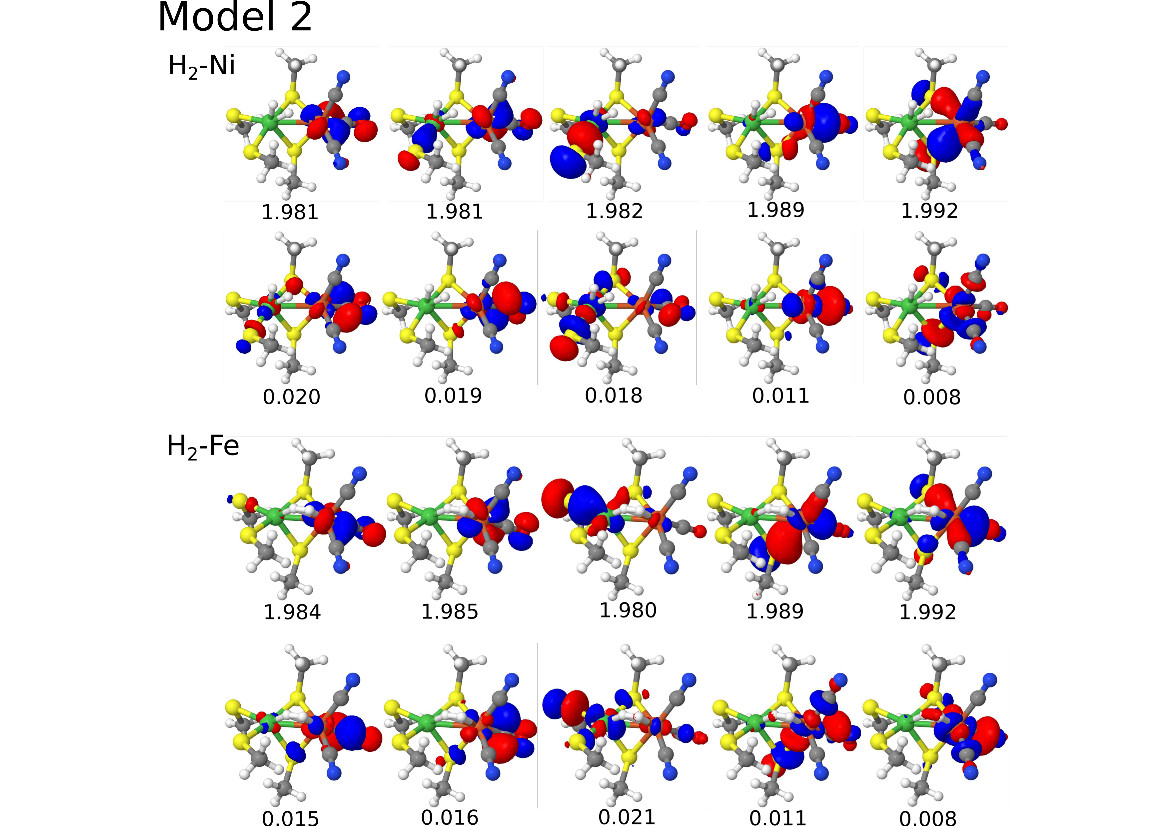}
\caption{ Active natural orbitals and their occupation numbers from a CAS(10,10)--srDFT calculation (started from the MP2 wavefunction) with $\mu=0.4$ for model 2\label{cas-10-10-orb2}} 
\end{figure} 

\begin{figure}
\centering
\includegraphics[width=1.0\textwidth]{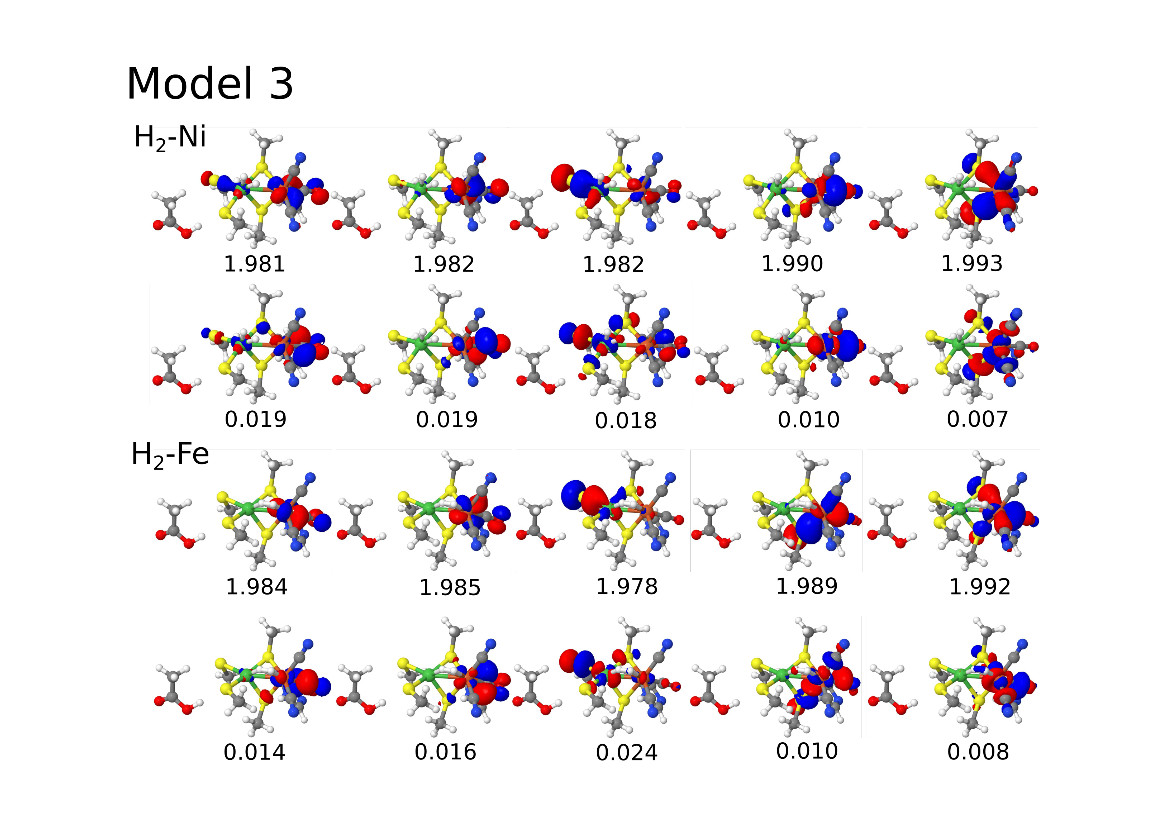}
\caption{ Active natural orbitals and their occupation numbers from a CAS(10,10)--srDFT calculation (started from the MP2 wavefunction) with $\mu=0.4$ for model 3\label{cas-10-10-orb3}} 
\end{figure}

\begin{figure}
\centering
\includegraphics[width=1.0\textwidth]{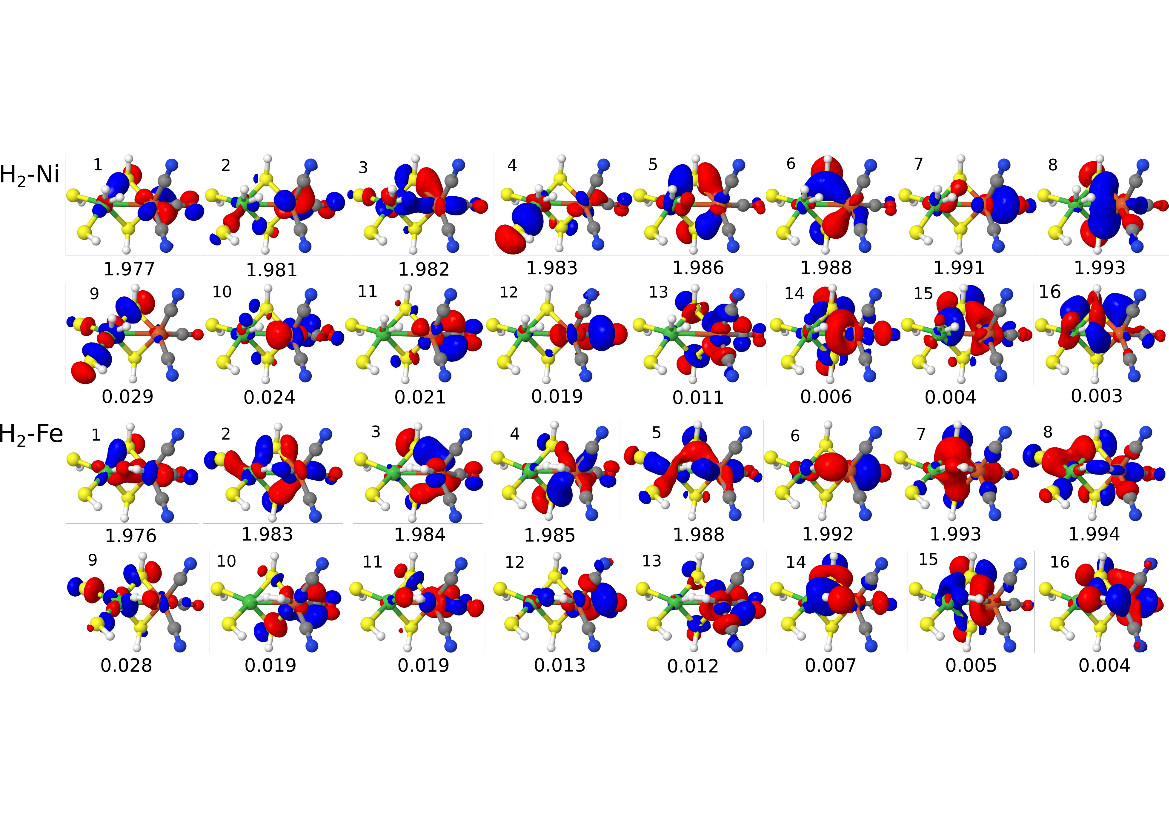}
\caption{Active natural orbitals and their occupation numbers from a CAS(16,16)--srDFT calculation for model 1 (started from the MP2 wavefunction) with $\mu=0.4$ \label{cas-16-16-orb}} 
\end{figure}



